\documentclass[12pt]{iopart}

\usepackage{iopams}
  
\usepackage{graphicx}

\begin{document}

\title[Exact solutions to the Lighthill Whitham Richards Payne traffic flow
equations]{Some exact solutions to the Lighthill Whitham Richards Payne traffic flow
equations}

\author{G Rowlands$^1$, E Infeld$^2$ and A A Skorupski$^2$}

\address{$^1$ Department of Physics, University of Warwick, Coventry CV4 7AL\\
$^2$ National Centre for Nuclear Research, Ho{\.z}a 69, 00--681 Warsaw, Poland
}

\eads{\mailto{g.rowlands@warwick.ac.uk}, \mailto{einfeld@fuw.edu.pl}, 
\mailto{askor@fuw.edu.pl}}

\begin{abstract}
We find a class of exact solutions to the Lighthill Whitham Richards Payne (LWRP)
traffic flow equations. Using two consecutive lagrangian transformations, a
linearization is achieved. Next, depending on the initial density, we either apply
(again two) Lambert functions and obtain exact formulas for the dependence of the car
density and velocity on $x,t$, or else, failing that, the same result in a parametric
representation. The calculation always involves two possible factorizations of a
consistency condition. Both must be considered. In physical terms, the lineup usually
separates into two offshoots at different velocities. Each velocity soon becomes
uniform. This outcome in many ways resembles the two soliton solution to the
Korteweg--de Vries equation. We check general conservation requirements. Although
traffic flow research has  developed tremendously since LWRP, this calculation, being
exact, may open the door to solving similar problems, such as gas dynamics or water
flow in rivers. With this possibility in mind, we outline the procedure in some detail
at the end.
\end{abstract}


\submitto{\JPA}

\section{General history. Formulation of the model}

We have found over the years that several nonlinear, partial differential equations
of physics, not integrable by standard methods, such as Inverse Scattering or else an
inversion of variables, yield their secrets to lagrangian coordinate methods
\cite{InfRol1}--\cite{SI}.
Here we will treat one such equation pair and see a combination of two
`lagrangian' transformations (the second one will be called quasi-lagrangian for
reasons to be explained) and a twofold introduction of the Lambert function enable us
to solve the one lane traffic flow problem explicitely. A further class of solutions
is found in parametric form. 

Although we find a class of solutions that falls short of being general, common sense
situations are well described, as well as some unexpected ones. The interesting
thing, however, is that the much researched nonlinear equations involved, known for
a long time now, can in some instances be solved exactly, without recourse to
approximations and with little or no numerics.

In 1955, James Lighthill and his former research student, Whitham, formulated an
equation describing single lane traffic flow, assumed congested enough to justify
a fluid model. The theory was formulated in the second part of their classic paper on
kinematic waves \cite{Light}.
Richards independently came to the same
conclusion and published in the following year \cite{Rich}. Next Payne
\cite{Payne} and Whitham \cite{Whith1} added a second equation and replaced the LWR
equation with standard continuity. We will call this pair LWRP. Recently the
literature on both models has grown considerably, see for example the books by Kern
\cite{Kern1} and further references \cite{Chandl}--\cite{Zhang}.

Extensions to more than one lane, lane changing, discrete models, higher order
effects, as well as numerical work, prevail. One of the original authors has found a
Toda lattice like solution to the discrete version of Newell \cite{New}, see
\cite{Whith2}. In a future paper, we will see if the methods introduced here can be
applied to some of these recent extensions of LWRP and LWR.

\subsection{The model}

Assume a long segment of a one lane road, deprived of entries and exits, sufficiently
congested by traffic and free of breakdowns to admit a continuous treatment, so as to
permit us to postulate the usual equation of continuity:
\begin{equation}
\frac{\partial\rho}{\partial t} + u \frac{\partial\rho}{\partial x} =
- \rho\frac{\partial u}{\partial x}.
\label{cont}
\end{equation}
Here $\rho$ is the density of cars, the maximum of which corresponds to a rather
deplorable, but all too familiar almost bumper to bumper situation, and $u$ is the
local velocity. The right hand side of the second, newtonian equation, formulated by
Payne \cite{Payne} and Whitham \cite{Whith1}, is less obvious:
\begin{equation}
\frac{\partial u}{\partial t} + u \frac{\partial u}{\partial x} =
\frac{V(\rho) - u}{\tau_0} - \frac{\nu_0}{\rho} \frac{\partial\rho}{\partial x} .
\label{mom}
\end{equation}
The first term on the right involves the mean drivers' reaction time $\tau_0$, and
the next term  models a diffusion effect depending on the drivers' awareness of
conditions beyond the preceding car. The constant $\nu_0$ is a sort of diffusion
coefficient, measuring the effect of the density gradient. Some recent improved models
also bring in the second derivative, but we will not at this stage.

In particular, $u = V(\rho)$ and $\rho$ both constant give a possible solution to the
LWRP equations. Indeed, in his book Whitham expands around this equilibrium and
treats the linear waves and possible instability so obtained \cite{Whith1}. However,
exact nonlinear solutions are what we are interested in at the present.

We specify
\begin{equation}
V(\rho) = V_0 - h_0 \tau_0 \sqrt{\nu_0} \rho, \qquad V_0 = \mathrm{const}.
\label{Vrho}
\end{equation}
This often postulated form of $V(\rho)$ is the only one that leads to an integrable
equation, as far as we can see. Fortunately, the plot of $Q(\rho) = \rho V(\rho)$,
which people measure, is always convex and parabola-like, see Whitham's figure 3.1.
Thus, so far luck is with us.

We introduce dimensionless variables by replacing
\begin{equation}
\fl t\ \to \ t\, \tau_0, \qquad (u,V_0) \to \ (u,V_0)\, \sqrt{\nu_0}, \qquad x \ \to
\ x\, \sqrt{\nu_0}\tau_0, \qquad \rho \to \ \rho\, (h_0\tau_0)^{-1}.
\label{reps}
\end{equation}
This leaves the continuity equation unchanged, and the newtonian equation takes the
form:
\begin{equation}
\frac{\partial u}{\partial t} + u \frac{\partial u}{\partial x} =
V_0 - \rho - u - \frac{1}{\rho} \frac{\partial\rho}{\partial x} .
\label{momds}
\end{equation}

\section{Introducing lagrangian coordinates}

The non-linearity on the left hand side of equations (\ref{cont}) and (\ref{momds})
can be eliminated by introducing lagrangian coordinates:
$\xi(x,t)$, the initial position (at $t=0$)
of a fluid element which at time $t$ was at $x$, and time $t$. In this description,
the independent variable $x$ becomes a function of $\xi$ and $t$, as are
the fluid parameters $\rho(\xi,t) = \rho(x(\xi,t),t)$ and $u(\xi,t) = u(x(\xi,t),t)$.

Here and in what follows we adopt the convention that a superposition of two functions
which introduces a new variable is denoted by the same symbol as the original
function, but of the new variable. Denoting by $f$ either $\rho$ or $u$, the basic
transformation between eulerian coordinates $x,t$ and lagrangian ones $\xi,t$ can be
written as
\begin{equation}\label{trans}
\fl x(\xi,t) = \xi + \int_0^t u(\xi,t') \, \mathrm{d}t', \qquad
\frac{\partial x}{\partial t} = u(\xi,t), \qquad \frac{\partial f(\xi,t)}{\partial t}
= \frac{\partial f(x,t)}{\partial t} + u \frac{\partial f(x,t)}{\partial x}.
\end{equation}
We denote by $s(\xi)$ the number of cars between the
last one at $\xi = \xi_{\mathrm{min}}$ and that at $\xi$:
\begin{equation}
s(\xi) = \int_{\xi_{\mathrm{min}}}^{\xi} \rho_0(\xi') \, \mathrm{d}\xi',
\qquad \rho_0(\xi) = \rho(x=\xi,t=0)
\label{sfxi}
\end{equation}
where $\rho_0(\xi)$ is the initial mass density distribution.

Here and in what follows, the subscript $0$ always refers to $t=0$. We will also use
the superscript $0$ to refer to $\xi=0$.

There are no gaps in the line of traffic considered and therefore $\rho_0(\xi)$
is positive.
Hence $s(\xi)$ is an increasing function starting at $s(\xi_{\mathrm{min}}) = 0$,
and one can introduce a uniquely defined inverse function $\xi(s)$. The initial
position of a fluid element can be specified by either $\xi$ or $s$.

If a small initial interval $\mathrm{d}\xi$ at $t=0$ becomes $\mathrm{d}x$
at time $t$, mass conservation requires:
\begin{equation}
\mathrm{d}s = \rho_0(\xi) \mathrm{d} \xi = \rho(x,t) \mathrm{d}x.
\label{mc}
\end{equation}
This leads to a mass conservation equation in lagrangian variables:
\begin{equation}\label{cont1}
\frac{\partial x(s,t)}{\partial s} = \frac{1}{\rho(s,t)},
\end{equation}
and to a useful operator identity
\begin{equation}\label{ss}
\frac{1}{\rho(x,t)}\frac{\partial}{\partial x} =
\frac{1}{\rho_0(\xi)}\frac{\partial}{\partial \xi} =
\frac{\partial}{\partial s}.
\end{equation}

Integrating (\ref{cont1}) over $s'$ from $s(\xi=0)$ to $s$, we obtain the continuity
equation in integral form:
\begin{equation}
X(s,t) \equiv x(s,t) - x(s^0,t) = \int_{s^0}^s \frac{\mathrm{d}s'}{\rho(s',t)},
\qquad s^0 = s(\xi=0).
\label{X}
\end{equation}
This indicates that if we know the car density in lagrangian coordinates $\rho(s,t)$,
we can determine the evolving shape of the line of traffic, where the distance $X$ is
measured from the $\xi=0$ car.

The analog of the continuity equation (\ref{cont}) is obtained by differentiating
(\ref{cont1}) by $t$. Using the middle part of (\ref{trans}) we obtain
\begin{equation}
\frac{\partial \psi(s,t)}{\partial t} = \frac{\partial u}{\partial s}, \qquad
\psi = \frac{1}{\rho},
\label{contlv}
\end{equation}

The newtonian equation in  lagrangian coordinates is obtained from (\ref{momds})
and (\ref{ss}):
\begin{equation}
\frac{\partial u(s,t)}{\partial t} + u = V_0 - \rho - \frac{\partial\rho}{\partial s}.
\label{mom1}
\end{equation}
Equation (\ref{mom1}) is linear and can be solved to express $u(s,t)$ in terms of
$\rho$. Again, using the middle part of (\ref{trans}) we can also calculate $x(s,t)$:
\begin{eqnarray}
u(s,t) &= \mathrm{e}^{-t} \biggl[\int_0^t N(s,t') \mathrm{e}^{t'} \, \mathrm{d}t' +
u(s,0) \biggr],
\label{uft}\\
N(s,t) &= V_0 - \Bigl[ \rho + \frac{\partial \rho}{\partial s} \Bigr],\label{Nst}\\
x(s,t) &= \xi(s) + \int_0^t u(s,t') \, \mathrm{d}t'\nonumber\\
&= \xi(s) + u(s,0) - u(s,t) + \int_0^t N(s,t') \, \mathrm{d}t',\label{xst}
\end{eqnarray}
where the function $u(s,0)$ will be determined later.

\section{Finding the fluid density}

Differentiating the newtonian equation (\ref{mom1}) by $s$, and using continuity
(\ref{contlv}), we obtain one equation for $\psi$:
\begin{equation}
\frac{\partial^2 \psi}{\partial t^2} - \frac{\partial}{\partial s} \Bigl( 
\frac{1}{\psi^2} \frac{\partial \psi}{\partial s} \Bigr) + 
\frac{\partial \psi}{\partial t} + \frac{\partial}{\partial s}
\frac{1}{\psi} = 0 .
\label{psieq}
\end{equation}
This equation can be factorized in two possible ways, I and II:
\begin{equation}
\mathrm{I:\ } \qquad\Bigl( \frac{\partial}{\partial t} + \frac{\partial}{\partial s}
\frac{1}{\psi} \Bigr) \Bigl( \frac{\partial \psi}{\partial t} -
\frac{1}{\psi} \frac{\partial \psi}{\partial s} + 1 + \psi \Bigr) = 0 ,
\label{fact1}
\end{equation}
and
\begin{equation}
\mathrm{II:} \qquad\Bigl( \frac{\partial}{\partial t} - \frac{\partial}{\partial s}
\frac{1}{\psi} \Bigr) \Bigl( \frac{\partial \psi}{\partial t} +
\frac{1}{\psi} \frac{\partial \psi}{\partial s} - 1 + \psi \Bigr) = 0 .
\label{fact2}
\end{equation}

We will find that the second factor in (\ref{fact1}) best yields solutions such that
$X \geq 0$, whereas that in (\ref{fact2}) rules $X<0$, where $X$ is always the
distance from the car that started at $x=0$. Different pairing would lead to trouble.

If we denote equation (\ref{psieq}) as
$O\psi = 0$ and the decompositions by I and II, the following symmetries hold: 
\begin{equation}
O = O_1^{\mathrm{I}}\, O_2^{\mathrm{I}} = O_1^{\mathrm{II}}\, O_2^{\mathrm{II}},
\qquad O_i^{\mathrm{I}}(\psi) = (-1)^{i+1} O_i^{\mathrm{II}}(-\psi), \qquad i=1,2 .
\end{equation}

In what follows, we will find solutions for which the second factor in one of
equations (\ref{fact1}), (\ref{fact2}) vanishes, leaving a more general treatment
to a possible later paper. We follow motion from left to right. Factorization also
means that we can only introduce the initial value of the density (or $\psi$).
The initial velocity $u(s,t=0)$ will then follow except for a universal constant.
We will have more to say about this later on.

The non-linearities in (\ref{fact1}) and (\ref{fact2}) (second factors) can be
eliminated if one transforms the variables $s,t$ to $\eta,t$ in a 
way similar to the lagrangian transformation (\ref{trans}),
though without the usual physical interpretation:
\begin{equation}\label{trans1}
\fl s(\eta,t) = \eta \mp \int_0^t  \frac{\mathrm{d}t'}{\psi(\eta,t')}, \qquad
\frac{\partial s}{\partial t} = \mp \frac{1}{\psi(\eta,t)}, \qquad
\frac{\partial \psi(\eta,t)}{\partial t} = \frac{\partial \psi(s,t)}{\partial t}
\mp \frac{1}{\psi} \frac{\partial \psi}{\partial s} \, .
\end{equation}
Solving the resulting linear equation
\[
\frac{\partial \psi(\eta,t)}{\partial t} = \mp 1 - \psi
\]
we obtain, in view of the fact that s and $\eta$ are identical at $t=0$,
\begin{equation}
\psi(\eta,t) = \mp 1 + \mathrm{e}^{-t} \bigl[ \psi_0(\eta) \pm 1 \bigr], \qquad
\psi_0(\eta) \equiv \psi(s=\eta,0).
\label{psia}
\end{equation}
For this $\psi(\eta,t)$ we have
\begin{equation}
\int_0^t \frac{\mathrm{d}t'}{\psi} = \mp \ln \bigl[ \mathrm{e}^t
\psi(\eta,t)/\psi_0(\eta) \bigr] ,
\label{inta}
\end{equation}
and finally, back to $\rho = 1/\psi$ and using (\ref{trans1}),
\begin{equation}
\fl s = \eta + \ln \bigl( 1 \mp \rho_0(\eta)A(t) \bigr), \qquad
\rho_0(\eta) = \rho_0(s=\eta), \qquad A(t) = \mathrm{e}^t - 1.
\label{sfeta}
\end{equation}
In this relation, defining $s$ in terms of $\eta$ and $t$, $\rho_0(\eta)$ is defined
by (\ref{sfxi}) but is expressed in terms $s$, where one has to rename $s$ to $\eta$.
Exactly the same procedure applies to $\psi_0(\eta)$ given by (\ref{psia}).

Using (\ref{psia}), we can express $\rho$ in terms of $\eta$ and $t$:
\begin{equation}
\rho(\eta,t) = \frac{1}{\mp 1 +\mathrm{e}^{-t}[1/\rho_0(\eta) \pm 1]},
\label{rhoa}
\end{equation}
which tends to $\rho_0(s)$ as $t \to 0$.

We are now in a position to determine the function $u(s,0)$ needed in equations
(\ref{uft}) and (\ref{xst}). Differentiate (\ref{mom1}) by $s$ and then subtract
both sides of (\ref{psieq}) from the result to obtain
\begin{equation}
\Bigl(\frac{\partial}{\partial t} + 1\Bigr)\Bigl(\frac{\partial\psi}{\partial t} -
\frac{\partial u}{\partial s}\Bigr)= 0.
\end{equation}
Solved by
\begin{equation}
\frac{\partial\psi}{\partial t} - \frac{\partial u}{\partial s} = f(s)e^{-t}.
\end{equation}
Therefore, if $f(s)=0$, equation (\ref{contlv}) will be valid for all time.
All we require is 
\begin{equation}
\Bigl[\frac{\partial\psi}{\partial t} - \frac{\partial u}{\partial s} 
\Bigr]_{t=0} = 0, \qquad \mathrm{i.e.} \qquad \frac{\partial u(s,0)}{\partial s} =
\frac{\partial \psi_0}{\partial t}.
\end{equation}
This result, along with either (\ref{fact1}) or (\ref{fact2}), leads to
\begin{equation*}
\frac{\partial u(s,0)}{\partial s} = \pm \frac{1}{\psi_0}
\frac{\partial \psi_0}{\partial s} \mp 1 - \psi_0.
\end{equation*}
Integrating over $s'$ from $s^0$ to $s$ and transforming the result to $\xi$, we
end up with
\begin{equation}
\fl u(\xi,0) = u_0 - \xi \mp \biggl[ s(\xi) - s^0 + \ln \frac{\rho_0(\xi)}{a} \biggr]
\qquad s^0 = s(\xi=0), \qquad a = \rho_0(\xi=0),
\label{us0}
\end{equation}
where $u_0 = u(\xi=0,0) \geq 0$ is arbitrary.

The last task is to determine $u(s,t)$, $x(s,t)$, and $X(s,t)$, given by (\ref{uft}),
(\ref{xst}), and (\ref{X}), in terms of $\eta$. Using (\ref{rhoa}), (\ref{sfeta}) and
(\ref{Nst}) we find the integrand $N$: 
\begin{eqnarray}
N(s,t) &= V_0 - \Bigl[ \rho + \frac{\partial \rho}{\partial s} \Bigr] =
V_0 - \Bigl[ \rho + \frac{\partial \rho/\partial \eta}{\partial s/\partial \eta}
\Bigr]\nonumber\\
&= V_0 \pm 1 - \frac{\pm 1 + \rho_0(\eta) + \rho_0'(\eta)}{1 \mp
[\rho_0(\eta) + \rho_0'(\eta)](\mathrm{e}^t - 1)},\label{Nstpar} 
\end{eqnarray}
which tends to $V_0 \pm 1$ as $t \to \infty$.
Here $\eta = \eta(s,t)$ must be found as a solution of the transcendental equation
(\ref{sfeta}), and the integrals (\ref{uft}) and (\ref{xst}) must be calculated
numerically. On the other hand, the integral (\ref{X}) can be calculated analytically:
\begin{eqnarray}
X(s,t) &\equiv x(s,t) - x(s^0,t) = \int_{\eta^0}^{\eta}
\frac{\partial s'/\partial \eta'}{\rho(\eta',t)} \, \mathrm{d} \eta'
\nonumber\\
&= \mathrm{e}^{-t} \biggl\{ \xi(s=\eta) - \xi(s=\eta^0)
\mp A(t) \biggl[ \eta - \eta^0 + \ln \frac{\rho_0(\eta)}{\rho_0(\eta^0)}
\biggr] \biggr\},
\label{Xpar}
\end{eqnarray}
where $\eta = \eta(s,t)$ and $\eta^0 = \eta(s^0,t)$ are defined implicitly by
(\ref{sfeta}).

Notice that the time evolution of an assumed initial density profile $\rho_0(\xi)$
can only be determined if the solution $\eta(s,t)$ of equation (\ref{sfeta}) is
a continuous function of $t$. This will be true either in case I, if $\rho_0(\xi)$
decreases from its initial value $a=\rho_0(\xi=0)$ as $\xi$ increases (i.e., for
$\xi \geq 0$), or in case II, if $\rho_0(\xi)$ grows as $\xi$ increases from
$-\infty$ to $0$.

In either case, $\eta(s,t)$ starts its time evolution with $\eta(s,0)=s$. Then in
case I, it grows to $\eta_{\mathrm{max}}(s) = \eta(s,t\to\infty)$, and in case II,
it falls to $\eta_{\mathrm{min}}(s) = \eta(s,t\to\infty)$.

If $\rho_0(\xi)$ has a maximum at $\xi = \xi_{\mathrm{m}}$ either for $\xi>0$ or
$\xi<0$, then for any $\xi$ within the interval $(0,\xi_{\mathrm{m}})$,
$\eta\bigl(s(\xi),t\bigr)$ as a function of $t$ would have a jump, which is
unacceptable.

Another obvious requirement on the initial density profile $\rho_0(\xi)$ is its
integrability over the interval $(-\infty,0)$ or $(0,\infty)$.

In practice, the general theory given in the last two sections is only useful if
the integral (\ref{sfxi}) can be determined analytically. Examples will be given
in the following sections.

\section{Two exponential profiles of the initial fluid density}

We will see that two exponential profiles of the initial fluid density
\begin{eqnarray}
\rho_0(\xi) = a \exp (-\lambda \xi), \qquad& \xi \geq 0, \qquad \mathrm{i.e.}
\qquad \xi_{\mathrm{min}}=0,\label{ic}\\
%
%
\rho_0(\xi) = a \exp (\lambda \xi), \qquad& \xi \leq 0, \qquad \mathrm{i.e.}
\qquad \xi_{\mathrm{min}} = - \infty,
\label{ic2}
\end{eqnarray}
play a special role here, as in their case it is possible to eliminate the auxiliary
variable $\eta$, and even find the fluid density $\rho$ in terms of $X$ and
$t$. This is because the argument of the logarithm in equation (\ref{sfeta}) defining
$\eta(s,t)$ is a linear function of $\eta$. Therefore the solution $\eta(s,t)$ of
(\ref{sfeta}) can be given in terms of the Lambert function $W(x)$ defined by
\begin{equation}
\fl W \exp(W) = x, \qquad \mathrm{equivalent\ to} \qquad W + \ln W = \ln x,
\qquad - 1/\mathrm{e} \leq x < \infty .
\label{W}
\end{equation}
The $W$ function is named after Johann Heinrich Lambert, Leonhard Euler's young
prot{\'e}g{\'e} and an important mathematician in his own right (1728--1777), see
\cite{Corl} for a summary. For negative $x$, $W$
has two negative branches and the upper one (continuous at $x=0$) along with the
main branch of the logarithms should be chosen in what follows ($\ln x =
\ln |x| +\mathrm{i}\pi$ when $x<0$).

The solution $Y$ of the transcendental equation
\begin{equation}
Y + \ln (aY + b) = c
\label{treq}
\end{equation}
is given by
\begin{equation}
Y = W\Bigl[ \exp \Bigl(c + \case{b}{a} - \ln a \Bigr) \Bigr] - \frac{b}{a}.
\label{solY}
\end{equation}

Using equation (\ref{sfxi}) we first find
\begin{equation}
s^0 = s(\xi=0) = \int_{\xi_{\mathrm{min}}}^0 \rho_0(\xi') \, \mathrm{d}\xi' =
\cases{
0  &for (\ref{ic}),\\
\frac{a}{\lambda}  &for (\ref{ic2}),}
\label{s0}
\end{equation}
and then calculate
\begin{eqnarray}
s(\xi) &= s^0 + \int_0^{\xi} \rho_0(\xi') \, \mathrm{d}\xi' =
s^0 \mp \frac{a}{\lambda} \Bigl( \exp (\mp \lambda \xi) - 1 \Bigr)\nonumber\\
&=\cases{
\frac{a}{\lambda} \Bigl( 1 - \exp (-\lambda \xi) \Bigr)
& for (\ref{ic}),\\
\frac{a}{\lambda} \exp (\lambda \xi) &for (\ref{ic2}).}
\label{s2c}
\end{eqnarray}
The inverse functions are given by
\begin{equation}
\xi(s) = \mp \frac{1}{\lambda} \ln \Bigl( 1 \mp \case{\lambda}{a} (s - s^0) \Bigr)
= \cases{
- \frac{1}{\lambda} \ln \Bigl( 1 - \case{\lambda s}{a} \Bigr)
&for (\ref{ic}),\\
\frac{1}{\lambda} \ln \frac{\lambda s}{a} &for
(\ref{ic2}).}
\label{xi2c}
\end{equation}
Using this formula we can transform the initial conditions (\ref{ic}) and (\ref{ic2})
given above in $x,t$ to $s,t$:
\begin{equation}
\rho_0(s) = a \mp \lambda (s - s^0)
= \cases{
a - \lambda s &for (\ref{ic}),\\
\lambda s &for (\ref{ic2}).}
\label{rho02c}
\end{equation}
We now look for solutions to equations (\ref{fact1}) and (\ref{fact2}) that recreate
the above initial conditions as $t$ tends to zero.

Replacing $s$ by $\eta$ in (\ref{rho02c}) and using the $\rho_0(\eta)$ so obtained in
(\ref{sfeta}), we can write the result as
\begin{equation}
s = \eta + \ln \Bigl\{ \lambda A(t)\,\eta + \Bigl[1 + A(t) ( \mp a - \lambda s^0 )
\Bigr] \Bigr\}.
\label{sfeta1}
\end{equation}
Indeed the argument of the logarithm is a linear function of $\eta$, and we can
use equations (\ref{treq}) and (\ref{solY}) to obtain
\begin{eqnarray}
\eta(s,t) &= w(s,t) \pm \frac{a}{\lambda} + s^0
-\frac{1}{\lambda A(t)}, \qquad A(t) = \mathrm{e}^t - 1,\nonumber\\
w(s,t) &= W \Bigl\{ \exp \Bigl[p(s,t)\Bigr] \Bigr\},
\label{etast1}\\
p(s,t) &= s - s^0 \mp \frac{a}{\lambda} + \frac{1}{\lambda A(t)} + \ln
\frac{1}{\lambda A(t)}.\nonumber
\end{eqnarray}

Inserting this $\eta(s,t)$ into (\ref{rhoa}) we obtain, in view of (\ref{rho02c})
\begin{equation}
\rho(s,t) = \pm \frac{\mathrm{e}^t}{A(t)} \Bigl[ \frac{1}{\lambda A(t) w(s,t)}
- 1 \Bigr].\label{rhost1}\\
\end{equation}
We recover the initial condition (\ref{rho02c}) when $t \to 0^+$, and so
$1/A(t) \to \infty$, by invoking the large argument approximation for W: $W(x)
\approx \ln(x) - \ln\ln(x)$, $x \gg 1$,
\begin{eqnarray}
\rho &\to \pm \frac{1}{A} \biggl[ \frac{1}{\lambda A \bigl(s - s^0 \mp
\frac{a}{\lambda} +
\frac{1}{\lambda A} \bigr)} - 1 \biggr] = \frac{a \mp \lambda (s-s^0)}{1 +
A(\lambda (s-s^0) \mp a))}\nonumber\\
&= a \mp \lambda (s-s^0) + \Or(A) .
\label{limrho}
\end{eqnarray}

Using (\ref{rhost1}) and (\ref{us0}) we can determine $N(\xi,t)$ and $u(\xi,0)$
needed in equations (\ref{uft})--(\ref{xst}), where $s = s(\xi)$ is given by
(\ref{s2c}):
\begin{eqnarray}
N(\xi,t) &= V_0 \pm 1 \pm \frac{\mathrm{e}^{-t}}{1 - \mathrm{e}^{-t}}
\biggl[ 1 - \frac{1}{\lambda (1 - \mathrm{e}^{-t}) \bigl(1 + w(s,t)\bigr)}
\biggr]\nonumber\\
&\to \quad V_0 \pm 1 \qquad \mathrm{as} \ t \to \infty,\label{Nst1}\\
u(\xi,0) &= u_0 + (\lambda - 1) \xi + \frac{a}{\lambda}
\Bigl[ \exp (\mp \lambda \xi) - 1 \Bigr],
\label{us01}
\end{eqnarray}
where in equation (\ref{Nst1}) we used the fact that for large $t$, $w(s,t) \approx
-\ln [\lambda(\mathrm{e}^t - 1)]$ which tends to minus infinity as $t \to \infty$.
Calculating the integrals in (\ref{uft}) and (\ref{xst}) numerically, we can find
characteristics $x(\xi,t)$ parametrized by the initial fluid element position $\xi$,
shown in figures 2 and 4.

We will now try to express $\rho$ directly in terms of $X$ by using (\ref{X}).
Inserting $\rho(s,t)$ given by (\ref{rhost1}) into (\ref{X}) and changing the
integration variable from $s$ to $w$ we obtain
\begin{eqnarray}
-\lambda X &= \pm \lambda^2 A^2(t) \mathrm{e}^{-t} \int_{s^0}^s \, \mathrm{d}s' \,
\frac{w(s',t)}{ \lambda A(t)w(s',t) - 1}\nonumber\\
&= \pm \lambda A(t) \mathrm{e}^{-t}
\int_{w(s^0,t)}^{w(s,t)} \mathrm{d}w' \,
\biggl[ 1 + \frac{1 + \lambda A(t)}{ \lambda A(t) w' - 1 } \biggr]\nonumber\\
&= \pm \mathrm{e}^{-t} \biggl\{\lambda A(t) \bigl[ w(s,t) - w(s^0,t) \bigr]
\nonumber\\
&\quad+
\bigl[ 1 + \lambda A(t) \bigr] \ln \frac{ \lambda A(t) w(s,t) - 1}{\lambda
A(t) w(s^0,t) - 1}
\biggr\},\label{Xcont}
\end{eqnarray}
which can be written as
\begin{equation*}
\fl Y + \ln \Bigl( \bigl[ 1 + \lambda A(t) \bigr] Y - 1 \Bigr) =
\frac{ \mp \lambda \mathrm{e}^t X + \lambda A(t) w(s^0,t)}{1 + \lambda A(t)}
+ \ln \bigl( \lambda A(t) w(s^0,t) - 1 \bigr),
\end{equation*}
where
\begin{equation*}
Y = \frac{\lambda A(t) w(s,t)}{1 + \lambda A(t)}.
\end{equation*}
Again using (\ref{treq}) and (\ref{solY}) to solve for $Y$, we find
$\lambda A(t) w(s,t)$ as a function of $X$ and $t$. Inserting it into (\ref{rhost1})
we end up with
\begin{equation}
\rho(X,t) = \cases{
\mp \frac{ [ 1 + \lambda A(t) ]
W\bigl(\mathrm{e}^{\phi}\bigr)}{(1 - \mathrm{e}^{-t})\{ 1 + [1 +
\lambda A(t)] W(\mathrm{e}^{\phi}) \}}, & $X \geq 0$,\\
0,  & $X < 0$,}
\label{rhoeul}
\end{equation}
where
\begin{equation}
\phi = \frac{ \mp \lambda \mathrm{e}^t X + \lambda A(t) w(s^0,t)
- 1}{1 + \lambda A(t)} + \ln \frac{ \lambda A(t) w(s^0,t) - 1}{1 + \lambda A(t)}.
\end{equation}
Note that equation (\ref{Xcont}) could also be obtained from (\ref{Xpar}), where
$\eta$ is given by (\ref{etast1}),
\[
\lambda \bigl[\xi(s=\eta) - \xi(s=\eta^0) \bigr] = \mp \ln
\frac{a \mp \lambda (\eta - s^0)}{a \mp \lambda (\eta^0 - s^0) }
\]
in view of (\ref{xi2c}), and $\rho_0$ is given by (\ref{rho02c}) with $s=\eta$.

As $t \to 0$ we recover the initial condition (\ref{ic}) or (\ref{ic2}) by once again
invoking the large argument approximation $W(x) \approx \ln(x) - \ln\ln(x)$. The
calculation is similar to that in $s,t$ above.

\subsection{Exponentially decreasing initial fluid density}

As initial condition at $t = 0$ we first take a pulse described by (\ref{ic})
and subsequently expected to move to the right. To the far right cars are
thin on the ground. We expect them to move freely and make it possible for the
congestion near the discontinuity to spread out. This common sense expectation will
test our result.

The density profile $\rho(X,t)$, defined by equation (\ref{rhoeul}) with upper sign
is shown in figure 1.
\begin{figure}
\centerline{\includegraphics[scale=0.6]{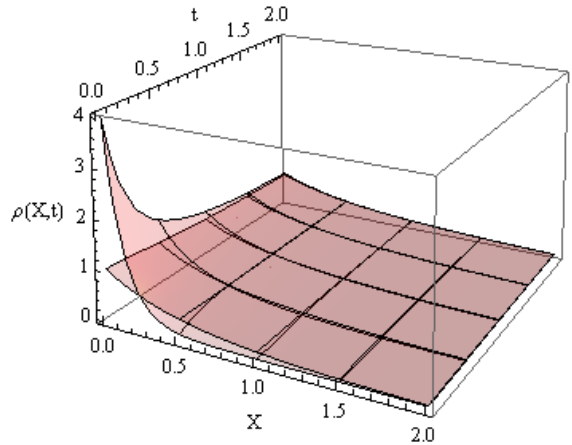}}
\caption{Two density profiles for various times as found
from our solution (I). Here $a =1$, $\lambda = 2$ in the first case, and $a = 4$,
$\lambda = 8$ in the second one. Nevertheless, the emerging  profiles are seen to be
identical after a while. The value of $a$ for each surface can be seen as equal to
$\rho(0,0)$.}
\end{figure}
\begin{figure}
\centerline{\includegraphics[scale=0.6]{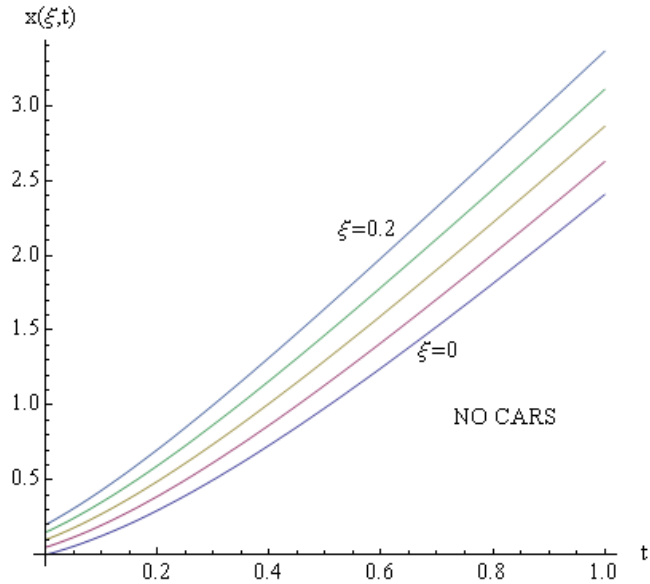}}
\caption{Characteristics $x(\xi,t)$ as functions of $t$
for $\xi$ ranging from $0$ to $0.2$ in steps of $0.05$, $a=4$, $\lambda = 8$,
$u_0 = 1$, and $V_0 = 2$.}
\end{figure}

As $t \to \infty$ this profile tends to
\begin{equation}
\rho \to \cases{
\frac{[ 1 - \exp(-a/\lambda) ] \exp(-X)}{1 -
[ 1 - \exp(-a/\lambda) ] \exp(-X)},& $X \geq 0$,\\
0, & $X < 0$.}
\label{rholim}
\end{equation}
That is because $A W \approx - [ 1 - \exp(-a/\lambda) ] \exp(-X)$, as small argument
$W$ functions can be approximated by their
argument, $W(x) \approx x$, regardless of sign.
The initial condition only appears as the ratio $a/\lambda$. A whole class of initial
profiles ends up developing identically, as long as this ratio is the same
(figure 1).

In figure 2 we present characteristics $x(\xi,t)$ obtained by numerically
calculating the integrals in (\ref{uft}) and (\ref{xst}). No crossing of
characteristic curves is observed, a pleasing vindication of the LWPR model.
In particular, the $\xi = 0$ line in figure 2 gives us the trajectory of the
discontinuity point, defining $X$.
\begin{figure}[t]
\centerline{\includegraphics[scale=0.6]{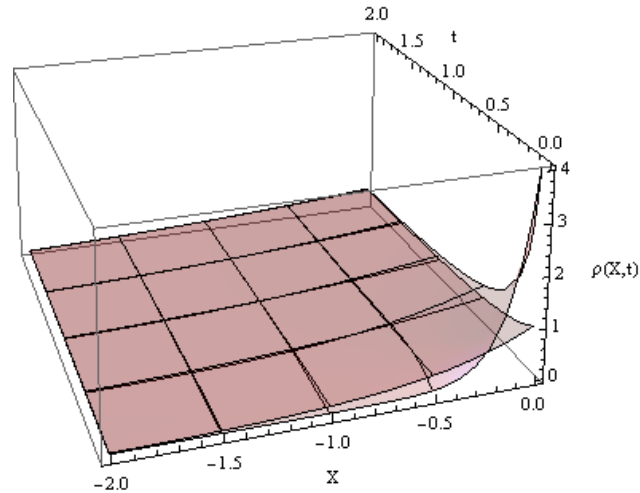}}
\caption{Two density profiles for various times as found
from our solution (II). Here $a =1$, $\lambda = 2$ in the first case, and $a = 4$,
$\lambda = 8$ in the second one. Nevertheless, the emerging  profiles are seen to be
identical after a while. The value of $a$ for each surface can be seen as equal to
$\rho(0,0)$. Notice that the final amplitude of $\rho$ is smaller here than in
figure 1.}
\end{figure}

Equation (\ref{Nst1}) with upper sign implies a uniform motion at the final, large
$t$ stage, at which $u = V_0 + 1$. This can also be seen from equation (\ref{mom}).
This emerging solution is identical to that given by Whitham \cite{Whith1} as a
special case of a `continuous shock structure', Whitham's equation (3.16), (when his
$A=0, U - v = 0$). Of course, continuity is lost at $s = 0$. Now we know exactly how
to set up initial conditions so as to obtain this profile. Alternatively, we could
set up this profile from the beginning. There are obviously fewer possible final
states than initial conditions, a state of affairs often encountered in non-linear
problems, see e.g. \cite{Armst,Jord}, and \cite{InfRol4} for further
references.

We can now check to see if the integral of $\rho$ is conserved in the $t \to \infty$
limit. In view of (\ref{rholim})
\begin{equation*}
\rho(X, t \to \infty) = \frac{\partial}{\partial X} \ln
\Bigl\{ 1 - [ 1 - \exp(-a/\lambda) ] \exp(-X) \Bigr\},
\label{rhoint}
\end{equation*}
which implies
\[
\int_0^{\infty} \rho(X, t \to \infty) \, \mathrm{d}X = a/\lambda \, .
\]
The integral over all $X$ of $\rho(X, t \to \infty)$ is indeed $a/\lambda$. Our exact
solution converges to a stationary mass conserving  travelling wave. 

\subsection{Exponentially increasing initial fluid density}

We now find a similar solution, but to case II. 
Our initial condition (\ref{ic2}) is now limited to $x \leq 0$.
 
This is the mirror image of the previous lineup, with the most congested traffic
facing an empty road. However, motion will still be from left to right.

The density profile $\rho(X,t)$, defined by equation (\ref{rhoeul}) with lower sign
is shown in figure 3. As $t \to \infty$ this profile tends to
\begin{equation}
\rho \to \cases{
\frac{[ \exp(a/\lambda) -1 ] \exp(X)}{1 +
[ \exp(a/\lambda) -1 ] \exp(X)},& $X \leq 0$,\\
0, & $X > 0$.}
\label{rholim2a}
\end{equation}

The end result is very similar to the previous one, but not identical.
Mass conservation follows from the same calculation. However, we are solving a
different equation and the asymptotic, uniform velocity is $V_0-1$ as opposed
to the $V_0+1$ of case I, see (\ref{Nst1}) with lower sign. This can also be seen
from figure 4, where characteristics $x(\xi,t)$ are presented.

\begin{figure}
\centerline{\includegraphics[scale=0.6]{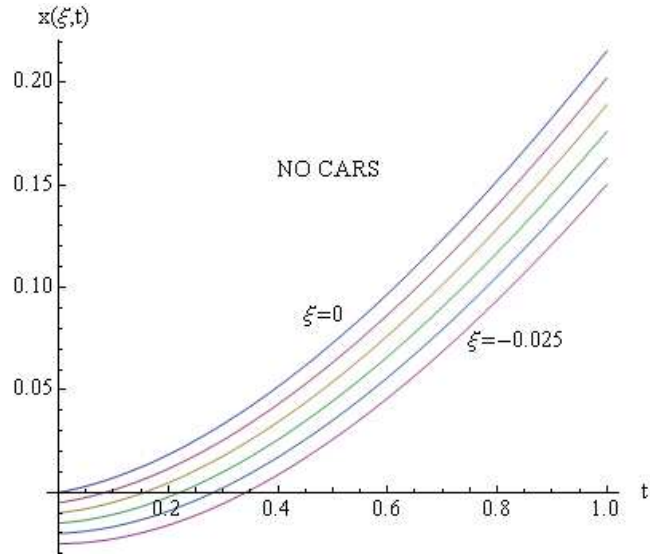}}
\caption{Characteristics $x(\xi,t)$ as functions of $t$
for $\xi$ ranging from $0$ to $-0.025$ in steps of $-0.005$, $a=1$, $\lambda = 2$,
$u_0 = 0.05$, and $V_0 = 10$.}
\end{figure}

If we match the initial velocities at the centre, $x=0$, then after a while, two
cavalcades emerge, a faster one at velocity $u = V_0 + 1$, and a slower one at
$V_0 - 1$ (cheaper cars?) see figure 4. Our situation reminds us of solutions to the
wave equation. Here $(c\,\partial_x - \partial_t)\psi = 0$ would correspond to I,
$(c\,\partial_x + \partial_t)\psi = 0$ to II. 
Coexistence of the two was not surprising in the linear world, now it is somewhat
unexpected. Although we are combining two solutions following from different
factorizations of the governing equation, we should remember that we are nevertheless
dealing with one exact solution, unique to the initial profile and velocities.

\section{The initial density profiles that can be treated parametrically}

In this section we present a few initial density profiles satisfying the
applicability conditions of our theory as formulated at the end of section 3,
see figure 5.
\begin{figure}
\centerline{\includegraphics[scale=0.6]{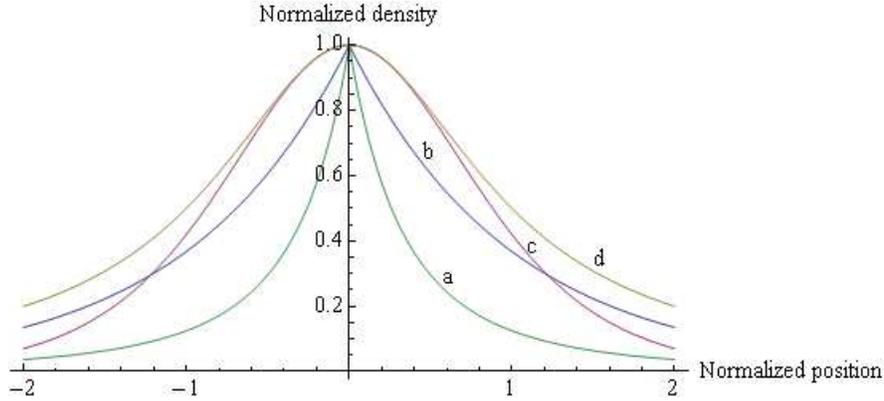}}
\caption{Normalized density profiles $\bar{\rho}_0=
\rho_0/a$ versus normalized position $\bar{\xi}=\lambda \xi$ for $\rho_0(\xi)$
given by (a): (\ref{ic}) and (\ref{ic2}), (b): (\ref{ic4}) for $b=1$,
$r=3$, (c): (\ref{icp}), and (d): (\ref{ic3}).}
\end{figure}

Detailed calculations will be performed for a pair of cases:
\begin{equation}
\rho_0(\xi) = \frac{a}{\cosh^2(\lambda \xi)} \equiv a \, \bigl[ 1 -
\tanh^2(\lambda \xi) \bigr],
\label{icp}
\end{equation}
where either $0 \leq \xi < \infty$ in case I, or $-\infty < \xi \leq 0$ in case II.

The fact that the derivative $\mathrm{d}\rho_0(\xi)/\mathrm{d}\xi$ vanishes
at $\xi=0$, in contrast to the exponential profiles (\ref{ic}) and (\ref{ic2}), will
have a consequence on the time evolution in case I, see figure 6.

The remaining profiles will have a power behaviour at infinity, $\rho_0(\xi)
\to (\pm\xi)^{-r}$ as $\pm\xi \to \infty$, where $r$ is a real number greater than
unity for integrability:
\begin{equation}
\rho_0(\xi) = \frac{a}{1 + (\lambda \xi)^2},
\label{ic3}
\end{equation}
and
\begin{equation}
\rho_0(\xi) = a \, \frac{b^r}{(\pm \lambda \xi + b)^r}, \qquad b > 0, \qquad r > 1,
\label{ic4}
\end{equation}
where the upper sign refers to case I, $\xi \geq 0$, and the lower one to case II,
$\xi \leq 0$.

By analogy to the exponential profiles (\ref{ic}) and (\ref{ic2}), each pair of
symmetric cases can be treated in a single calculation. For $\rho_0(\xi)$ given by
(\ref{icp}) we first find
\begin{equation}
s^0 = s(\xi=0) = \int_{\xi_{\mathrm{min}}}^0 \rho_0(\xi') \, \mathrm{d}\xi' =
\cases{
0 & for $\xi \geq 0$,\\
\frac{a}{\lambda} & for $\xi \leq 0$,}
\label{s0par}
\end{equation}
and then calculate
\begin{equation}
s(\xi) = s^0 + \int_0^{\xi} \rho_0(\xi') \, \mathrm{d}\xi' =
s^0 + \frac{a}{\lambda} \tanh(\lambda \xi).\label{sfxip}
\end{equation}
The inverse functions are given by
\begin{equation}
\xi(s) = \frac{1}{2\lambda} \ln
\frac{1 + \lambda (s - s^0)/a}{1 - \lambda (s - s^0)/a}
= \cases{
\frac{1}{2\lambda} \ln \frac{1 + \lambda s/a}{1 - \lambda s/a}
& for $\xi \geq 0$,\\
\frac{1}{2\lambda} \ln \frac{\lambda s/a}{2 - \lambda s/a}
& for $\xi \leq 0$.}
\label{xifsp}
\end{equation}
Using $\tanh(\lambda \xi)$ calculated from (\ref{sfxip}) in (\ref{icp}) we obtain
\begin{equation}
\rho_0(s) = a \Bigl[ 1 - \Bigl( \lambda (s - s^0)/a \Bigr)^2 \Bigr]
= \cases{
a [ 1 - (\lambda s/a)^2 ] & for $\xi \geq 0$,\\
\lambda s (2 - \lambda s/a) & for $\xi \leq 0$.}
\label{rhosp}
\end{equation}
Replacing here $s$ by $\eta$ and using the $\rho_0(\eta)$ so obtained in
(\ref{sfeta}) and (\ref{Nstpar}) along with (\ref{sfxip}) we find equations
defining $\eta(\xi,t)$ and the integrand $N(\eta,t)$ needed in equations
(\ref{uft})--(\ref{xst}):
\begin{eqnarray}
\frac{a}{\lambda} \tanh(\lambda \xi) = \eta + \ln
\Bigl( 1 - a [1 - (\lambda\eta/a)^2] A(t) \Bigr), \qquad& \mathrm{for}
\ \xi \geq 0,\label{etaeq1}\\
\frac{a}{\lambda} \bigl[ 1 + \tanh(\lambda \xi) \bigr] = 
\eta + \ln\Bigl( 1 + \lambda \, \eta (2 - \lambda\eta/a) A(t) \Bigr),
\qquad& \mathrm{for} \ \xi \leq 0,\label{etaeq2}
\end{eqnarray}
\begin{equation}
N(\eta,t) = V_0 \pm 1 + \frac{\mp 1 - f(\eta)}{1 \mp f(\eta)A(t)},\label{Netat}
\end{equation}
where
\begin{equation}
f(\eta) =
\cases{
- \eta (\eta + 2)\lambda^2/a & for $\xi \geq 0$,\\
\lambda \Bigl[ -\eta^2\lambda/a + 2\eta (1 - \lambda/a)
+ 2 \Bigr] & for $\xi \leq 0$.}
\end{equation}

In a similar way we can determine $X(\eta,t)$ by using (\ref{Xpar}) along with
(\ref{xifsp}) and (\ref{rhosp}) with $s=\eta$:
\begin{eqnarray}
X &= - \frac{\mathrm{e}^{-t}}{\lambda} \biggl[\Bigl( \lambda A(t) + \frac{1}{2}
\Bigr) \ln
\frac{1 - \lambda\eta/a}{1 - \lambda\eta^0/a} + \Bigl( \lambda A(t) - \frac{1}{2}
\Bigr) \ln
\frac{1 + \lambda\eta/a}{1 + \lambda\eta^0/a}\nonumber\\
&\quad + \lambda A(t)(\eta - \eta^0) \biggr] \qquad \mathrm{for} \ \xi \geq 0,
\label{X1p}
\end{eqnarray}
\begin{eqnarray}
X &= \frac{\mathrm{e}^{-t}}{\lambda} \biggl[\Bigl( \lambda A(t) + \frac{1}{2}\Bigr)
\ln\frac{\eta}{\eta^0} + \Bigl( \lambda A(t) - \frac{1}{2} \Bigr)
\ln \frac{2 - \lambda\eta/a}{2 - \lambda\eta^0/a} \nonumber\\
&\quad + \lambda A(t)
(\eta - \eta^0) \biggr] \qquad \mathrm{for} \ \xi \leq 0.\label{X2p}
\end{eqnarray}

Using here  $\eta(\xi,t)$ defined implicitly by either of equations (\ref{etaeq2})
and in $\rho(\eta,t)$ given by (\ref{rhoa}), we obtain $\rho(X,t)$ in parametric form:
$\rho(\xi,t)$ and $X(\xi,t)$. This form is appropriate for making use of the
ParametricPlot3D command of \textit{Mathematica}. The results are shown in
figures 6 and 7. They resemble those shown in figures 1 and 2 except for the time
evolution of the discontinuity at $X=0$, $\rho(X=0,t)$ for case I ($\xi\geq 0$),
shown in figure 6.
\begin{figure}
\centerline{\includegraphics[scale=0.6]{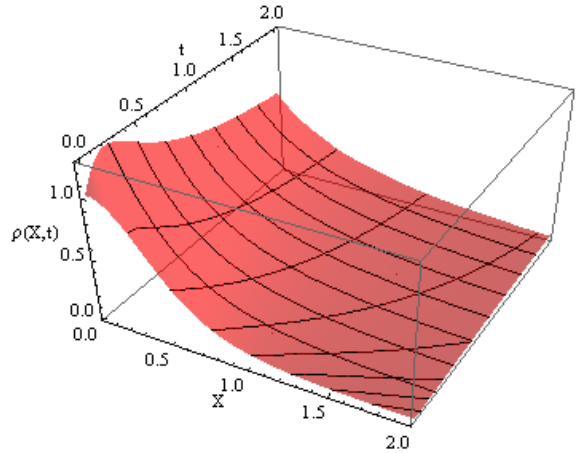}}
\caption{The fluid density represented parametrically
as found from our solution (I). Here $a =1$, $\lambda = 2$. The mesh lines correspond
to $t=\mathrm{const}$ or $\xi = 0,\:0.25,\:0.5,\:0.75, \dots $. Each value of $\xi$
is equal to $X$ at $t=0$.}
\end{figure}
\begin{figure}[t]
\centerline{\includegraphics[scale=0.7]{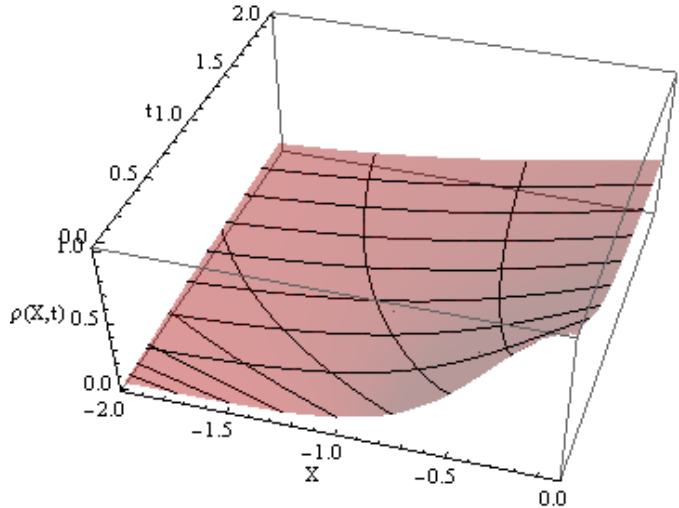}}
\caption{The fluid density as in figure 6 but as found
from our solution (II). Here again $a =1$, $\lambda = 2$, and $\xi = 0,\:-0.25,\:
-0.5,\:-0.75,\dots $, see $X$ at $t=0$.}
\end{figure}

The characteristics $x(\xi,t)$ can be found from equations (\ref{uft})--(\ref{xst}) by
numerical integration, where the integrand $N(\xi,t)$ is defined by (\ref{Netat})
and either of equations (\ref{etaeq2}), and
\begin{equation}
u(\xi,0) = s_0 - \xi \mp \Bigl\{ \frac{a}{\lambda} \tanh(\lambda \xi) - 2 \,
\ln \Bigl[ \cosh(\lambda \xi) \Bigr] \Bigr\},\label{uxi0}
\end{equation}
see equations (\ref{us0}, (\ref{icp}) and (\ref{sfxip}). The results, depending
on two parameters $V_0$ and $u_0$, are shown in figure 8.
\begin{figure}
\centerline{\includegraphics[scale=0.7]{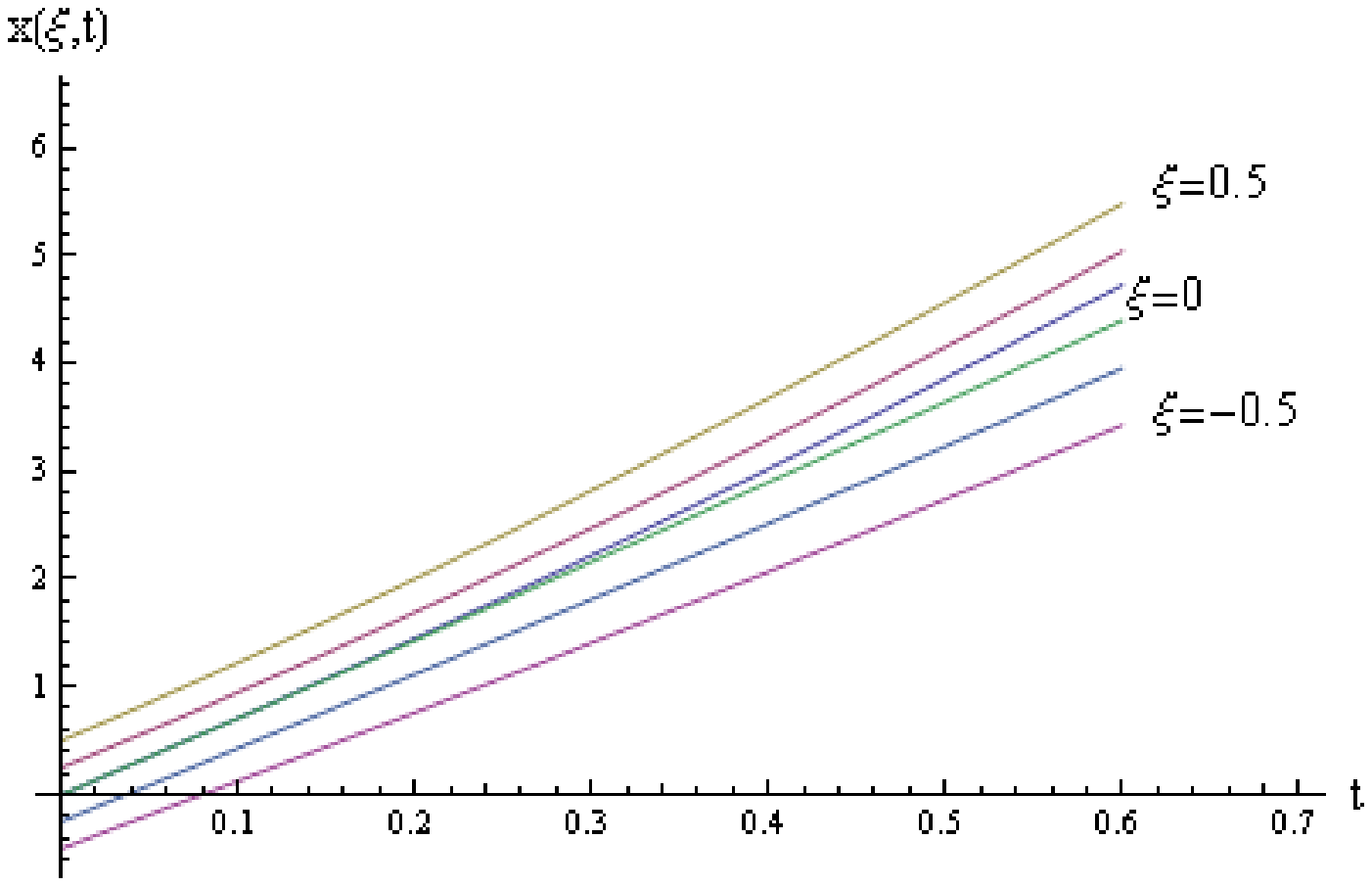}}
\caption{Characteristics $x(\xi,t)$ as functions of $t$
for $\xi=0,\: 0.25,\: 05,$ and $\xi=0,\: -0.25,\: -05$, $a=1$, $\lambda = 2$,
$u_0 = 7$, and $V_0 = 10$.}
\end{figure}

A characteristic feature of the plots representing the density given in parametric
form, $\rho(\xi,t)$ and $X(\xi,t)$, is that the mesh lines correspond to
$\xi = \mathrm{const}$, and $t = \mathrm{const}$, see figures 6 and 7. For the density
given explicitly, $\rho(X,t)$, they correspond to $X = \mathrm{const}$, and
$t = \mathrm{const}$, see figures 1 and 2. The mesh lines $\xi = \mathrm{const}$ are
particularly useful. Each point on a $\xi = \mathrm{const}$ mesh line gives us the
information of both the actual position $X$ and the associated density at time $t$,
for the car that started from $X = \xi$ at $t=0$. This information is given in the
frame moving along with the discontinuity at $\xi=0$. The motion of these frames
in turn is described by the characteristics labelled with $\xi=0$ in figure 8.

Adding cases I and II, we have a solution such that the initial congestion splits in
the middle, resulting once again in a slower cavalcade following a faster one, see
figure 8. This is rather like a two soliton solution of the Korteweg--de Vries
equation, see e.g. \cite{InfRol4}.

\section{Summary}

Our solutions augment those found for simpler, single equation nonlinear models,
e.g. Burgers, see \cite{Whith1}. Our exact solutions converge to single or double
stationary travelling wave structures after a few $\tau_0$ (figure 1). This steady
state convergence is more or less what one would expect, at least in the single case.
The LWRP equations have several families of solutions, to which we hereby add.
Generally we must use common sense to home in on the relevant ones. However, exact
solutions help and absolve us from the need to guess. As in police work, connecting
initial conditions with a present situation requires painstakingly following through
the history. Exact solutions exempt us of this task.

We introduce a lagrangian transformation in $x,t$. The variables are now $\xi,t$,
$\xi$ being the position at $t=0$. It labels the car rather than its actual position.
The continuity equation is now linear, simplifying further calculations. It may so
happen that the newtonian equation can yield a condition in just one dependent
variable. If this condition can be factorized, there is a possibility of finding exact
solutions by finding solutions to this factorized equation, being of lower order.
In our case, two distinct factorizations were possible, both were solved, and both
solutions were needed to cover all $x$. In fact, each factorization was assigned to
a specific half axis, the same for all cases considered. This situation, demanding
merging two solutions to cover the whole domain, is unusual in a nonlinear problem.
Another requirement was that the original profile have only one maximum at $x=0$.

We had to introduce a second, quasi lagrangian transformation to linearise. In
general, we obtain the solution in parametric form. However, if we are lucky, we can
solve completely, that is get rid of the parameter, by using Lambert functions. We
give an example of this.

It  should be stressed that a complete solution is only possible if we combine our
two factorized equations, I and II. This is similar to the situation for solutions
to the wave equation
\begin{equation*}
\Bigl(c^2 \frac{\partial^2}{\partial x^2} - \frac{\partial^2}{\partial t^2}\Bigr)
\psi = 0
\end{equation*}
such that solutions of $(c\,\partial_x - \partial_t)\psi = 0$ and
$(c\,\partial_x + \partial_t)\psi = 0$ coexist. However, this coexistence is somewhat
surprising for a nonlinear system.

An interesting question is: how wide a class of non-linear problems can be so solved?
Perhaps one possibility is furnished by the somewhat similar gas dynamics and shallow
water equations? In our case, exponential initial conditions alone seem to promise
that the Lambert function will make an appearance. Perhaps they should be used for
similar problems.

Another interesting question is how general is the requirement of merging solutions
I and II when two factorizations of the consistency condition exist.

\ack{We thank Drs M. Grundland and P. Goldstein for pointing out that our equations
I and II are both integrable and non-Painlev\'e, a rare combination in mathematical
physics.}


\section*{References}

\begin{thebibliography}{99}

\bibitem {InfRol1} Infeld E and Rowlands G 1989 Relativistic bursts \PRL \textbf{62}
1122--25

\nonum Infeld E and Rowlands G 1997 Lagrangian picture  of plasma physics
I \textit{J. Tech. Phys.} \textbf{38} 607--45

\nonum Infeld E and Rowlands G 1998 Lagrangian picture  of plasma physics
II \textit{J. Tech. Phys.} \textbf{39} 3--35

\bibitem {InfRol4} Infeld E and Rowlands G 2000 \textit{Nonlinear Waves, Solitons and
 Chaos} 2nd edn (Cambridge: Cambridge University Press)

\bibitem {IRS} Infeld E, Rowlands G and Skorupski A A 2009 Analytically solvable
model of nonlinear oscillations in a cold but viscous and resistive plasma
\PRL \textbf{102} 145005
(doi: 10.1103/PhysRevLett.102.145005)

\bibitem {SI} Skorupski A A and Infeld E 2010 Nonlinear electron oscillations in a
viscous and resistive plasma
\textit{Phys. Rev.} E \textbf{81} 056406
(doi: 10.1103/PhysRevE.81.056406)

\bibitem {Light} Lighthill M J and Whitham G B 1955 On kinematic waves, I Fluid
movement in long rivers; II Theory of traffic flow on long crowded roads
\PRS A \textbf{229} 281--345
(doi: 10.1098/rspa.1955.0088)

\bibitem {Rich} Richards P I 1956 Shock waves on the highway \textit{Oper. Res.}
\textbf{4} 42--51
 
\bibitem {Payne} Payne H J 1971
\textit{Mathematical models of public systems in simulation Councils Proceedings}
\textbf{1} 51--60

\bibitem {Whith1} Whitham G B 1974 \textit{Linear and Nonlinear Waves} chap 3
(New York: John Wiley)

\bibitem {Kern1} Kern B S 2003 \textit{The physics of traffic flow} (Berlin: Springer)

\nonum Kern B S 2009 \textit{Introduction to Modern Traffic Flow, Theory and
Control} (Berlin: Springer) 

\bibitem {Chandl} Chandler R E, Herman R and Montroll E W 1958 Traffic dynamics:
studies in car following \textit{Oper. Res.} \textbf{6} 165--84

\bibitem {Greenb} Greenberg  H 1959 An analysis of traffic flow
\textit{Oper. Res.} \textbf{7} 79--85

\bibitem {Herman} Herman R, Montroll E W, Potts R B and Rothery R W 1959 Traffic
dynamics: analysis of stability in car following
\textit{Oper. Res.} \textbf{7} 86--106
 
\bibitem {Jin} Jin W L and Zhang H M 2003 The formation and structure of vehicle
clusters in the Payne--Whitham traffic flow model \textit{Transportation Research}
\textbf{37} 207--23
 
\bibitem {Kerner} Kerner B S, Klenov L and Konhauser P 1997 Asymptotic theory of
traffic flow \textit{Phys. Rev.} E \textbf{56} 4200--16

\bibitem {Koment} Komentani E and Sasaki T 1958 On the stability of traffic flow
\textit{Oper. Res. Jpn.} \textbf{2} 11--26
 
\bibitem {Lua} Lua Y \etal 2008 Explicit construction of solutions for the
Lighthill--Whitham--Richards traffic flow model with a piecewise traffic flow
density model \textit{Transportation Research} \textbf{42} 355--72

\bibitem {Papa} Papageorgiou M 1983 \textit{Application of Automatic Control Concepts
in Traffic Flow and Control} (Berlin: Springer)

\bibitem {Zhang} Zhang M, Shu C, Wong G C K and Wong S C 2003 A weighted
essentially non-oscillatory numerical scheme for a multi-class LWR traffic flow
model \textit{J. Comp. Phys.} \textbf{191} 639--59

\bibitem {New} Newell G F 1961 Nonlinear effects in the dynamics of car following
\textit{Oper. Res.} \textbf{9} 209--29

\bibitem {Whith2} Whitham G B 1990 Exact solutions for a discrete system arising in
traffic flow \PRS A \textbf{428} 49--69
(doi:10.1098/rspa.1990.0025 )

\bibitem {Corl} Corless R M, Gonnet G H, Hare D E G, Jeffrey D J and
Knoth D E 1996 On the Lambert $W$ function \textit{Adv. in Comp. Math.} \textbf{5}
329--59

\bibitem {Armst} Armstrong T and Montgomery D 1967 Asymptotic state of the two-stream
instability \textit{J. Plasma Phys.} \textbf{1} 425--33
(doi: 10.1017/S0022377800003421)

\bibitem {Jord} Jordan P M and Puri P 2002 Exact solution for the unsteady plane
Couette flow of a dipolar fluid \PRS A \textbf{458} 1245--72

\end{thebibliography}
\end{document}